\def\pr#1 {Phys. Rev. {\bf D#1\tie \rm }}
\def\pe#1 {Phys. Rev. {\bf #1\tie\rm }}
\def\pre#1 {Phys. Rep. {\bf #1\tie\rm }}
\def\pl#1 {Phys. Lett. {\bf #1B\tie \rm }}
\def\prl#1 {Phys. Rev. Lett. {\bf #1\tie \rm }}
\def\np#1 {Nucl. Phys. {\bf B#1\tie \rm }}
\def\ap#1 {Ann. Phys. (NY) {\bf #1\tie \rm }}
\def\cmp#1 {Commun. Math. Phys. {\bf #1\tie \rm }}
\def\imp#1 {Int. Jour. Mod. Phys. {\bf A#1\tie \rm }}
\def\mpl#1 {Mod. Phys. Lett. {\bf A#1\tie\rm }}
\def\jhep#1 {JHEP {\bf #1\tie\rm }}
\def\zp#1 {Z. Phys. {\bf C#1\tie\rm }}

\def\tie{\noexpand~}
\def\ov{\overline}
\def\s{(\sigma)}

\def\sp{(\sigma ')}

\def\Tb{\overline T}
\def\d{\delta(\sigma-\sigma ')}
\def\dpr{\delta^{\,\prime}(\sigma-\sigma ')}
\def\dppr{\delta^{\,\prime\prime}(\sigma-\sigma ')}
\def\dpppr{\delta^{\,\prime\prime\prime}(\sigma-\sigma ')}

\def\ints{\int d\sigma\,}

\def\dx{\partial X}
\def\dbx{{\overline\partial}X}

\newcommand{\I}{{\rm i}}
\def\ra{\rightarrow}
\def\be{\begin{equation}}
\def\ee{\end{equation}}
\def\bea{\begin{eqnarray}}
\def\eea{\end{eqnarray}}
\def\Ref#1{(\ref{#1})}

\def\nn{\nonumber}

\catcode`@=11
\def\marginnote#1{}
\newcount\hour
\newcount\minute
\newtoks\amorpm
\hour=\time\divide\hour by60
\minute=\time{\multiply\hour by60 \global\advance\minute by-\hour}
\edef\standardtime{{\ifnum\hour<12 \global\amorpm={am}%
        \else\global\amorpm={pm}\advance\hour by-12 \fi
        \ifnum\hour=0 \hour=12 \fi
        \number\hour:\ifnum\minute<10 0\fi\number\minute\the\amorpm}}
\edef\militarytime{\number\hour:\ifnum\minute<10 0\fi\number\minute}
\def\draftlabel#1{{\@bsphack\if@filesw {\let\thepage\relax
   \xdef\@gtempa{\write\@auxout{\string
      \newlabel{#1}{{\@currentlabel}{\thepage}}}}}\@gtempa
   \if@nobreak \ifvmode\nobreak\fi\fi\fi\@esphack}
        \gdef\@eqnlabel{#1}}
\def\@eqnlabel{}
\def\@vacuum{}
\def\draftmarginnote#1{\marginpar{\raggedright\scriptsize\tt#1}}
\def\draft{\oddsidemargin 0.0truein
        \def\@oddfoot{\sl preliminary draft \hfil
        \rm\thepage\hfil\sl\today\quad\militarytime}
        \let\@evenfoot\@oddfoot \overfullrule 3pt
        \let\label=\draftlabel
        \let\marginnote=\draftmarginnote
   \def\@eqnnum{(\theequation)\rlap{\kern\marginparsep\tt\@eqnlabel}%
\global\let\@eqnlabel\@vacuum}  }
\catcode`@=12
\documentstyle[epsf, 12pt]{article}
\begin{document}

\thispagestyle{empty}
\begin{flushright}
RU01-06-B
\end{flushright}

\bigskip\bigskip
\begin{center}
\Large{\bf SPACETIME SUPERSYMMETRY IN A\\[2mm]
NONTRIVIAL NS-NS SUPERSTRING\\[2mm]
BACKGROUND}
\end{center}

\vskip 1.0truecm

\centerline{\bf Jonathan Bagger\footnote{bagger@jhu.edu}}

\vskip5mm
\centerline{\it Department of Physics and Astronomy}
\centerline{\it Johns Hopkins University}
\centerline{\it 3400 North Charles Street}
\centerline{\it Baltimore, MD 21218}
\vskip5mm

\centerline{\bf Ioannis
Giannakis\footnote{giannak@theory.rockefeller.edu}}
\vskip5mm
\centerline{\it Physics Department}
\centerline{\it Rockefeller University}
\centerline{\it 1230 York Avenue}
\centerline{\it New York, NY 10021}

\vskip5mm

\bigskip \nopagebreak \begin{abstract}
\noindent
In this paper we consider superstring propagation in
a nontrivial NS-NS background.  We deform the world
sheet stress tensor and supercurrent with an infinitesimal
$B_{\mu\nu}$ field.  We construct the gauge-covariant
super-Poincar\'e generators in this background and show that the
$B_{\mu\nu}$ field spontaneously breaks spacetime supersymmetry.
We find that the gauge-covariant spacetime momenta cease to
commute with each other and with the spacetime supercharges.
We construct a set of ``magnetic" super-Poincar\'e generators
that are conserved for constant field strength $H_{\mu\nu\lambda}$,
and show that these generators obey a ``magnetic" extension
of the ordinary supersymmetry algebra.
\end{abstract}

\newpage\setcounter{page}1

\vfill\vfill\break

\section{Introduction}

Two-dimensional superconformal field theories are solutions
to the classical superstring equations of motion.  Their
infinitesimal deformations \cite{nontrivial} can be used
to study superstring propagation in nontrivial backgrounds
\cite{solutions} and to elucidate the symmetry structure
of string theory itself \cite{symmetry}.

In this paper we describe superstring propagation in a nontrivial
NS-NS background.  We start in section 2 by deriving the
infinitesimal deformations that preserve the superconformal
structure.  We show that they also preserve the nilpotency of
the BRST operators.  We then construct the deformation that
describes superstring propagation in a nontrivial two-form
NS-NS background.

In sections 3 and 4 we use this formalism to study superstring
propagation in the two-form background.  We construct the
gauge-covariant super-Poincar\'e generators and compute the
spacetime supersymmetry algebra in the presence of the two-form
field.  We find that the supersymmetry is spontaneously broken,
and that the gauge-covariant spacetime momenta cease to
commute with each other and with the spacetime supercharges.

In section 5 we restrict our attention to the case of a
constant three-form field strength.  We construct a set
of conserved ``magnetic" super-Poincar\'e generators that
give rise to a ``magnetic" extension of the supersymmetry
algebra.  The magnetic supersymmetry is a generalization
of the magnetic translational symmetry associated with
point particles in a constant magnetic field \cite{hall}.

We conclude with an Appendix in which we derive the most
general two-form deformation that preserves the superconformal
structure.

\section{Superconformal Deformations}

In this paper we work in a Hamiltonian formalism
in which the two-dimensional world sheet is parametrized
by variables $\sigma$ and $\tau$.  We define our
superconformal field theory by three elements: i)~an
algebra of operators, $\cal A$,  ii)~a representation
of that algebra, and iii)~two distinguished elements
of $\cal A$, the holomorphic and antiholomorphic stress
energy superfields ${\cal T}(\sigma,\theta) = T_F\s +
\theta T\s$ and ${\overline{\cal T}}(\sigma,\theta) =
{\overline T}_F\s + {\overline{\theta}}{\overline T}\s$.
The holomorphic components satisfy one copy of the
super-Virasoro algebra,
\bea
[ T\s, T\sp ] &=& -\,{\I c \over 24\pi}\,\dpppr
\ +\ 2 \I \,T\sp\,\dpr\, -\,  \I\, T'\sp\,\d \nn\\[2mm]
\left\{ T_F\s, T_F\sp \right\} &=& -\,{1\over 2 \sqrt 2}\,T\sp\,\d\, +\,
{c\over 24 \sqrt 2 \pi}\,\dppr \nn\\[2mm]
[ T\s, T_F\sp ] &=& {3\I\over 2}\,T_F\sp\,\dpr\, -\, \I\,
T^{\prime}_F\sp\,\d\ ;
\eea
the antiholomorphic components satisfy another.
The operators $T\s$ and ${\overline T}\s$ are the
bosonic stress energy tensors, while $T_F\s$ and
${\overline T}_F\s$ are their supersymmetric partners.

The algebra $\cal A$ includes superfields $\Phi\s$ with
bosonic and fermionic components, $\Phi\s = \Phi_B\s +
\theta \Phi_F\s$.  It also includes spin fields $S^{\alpha}
\s$ whose presence renders $\cal A$ nonlocal.  The
states of the theory span representations of the
super-Virasoro algebra. The highest weight states are created
by superprimary fields, defined to be superfields
$\Phi\s$ whose components satisfy
\bea
[T\s, {\Phi_F}\sp]&=&
\I\, d\, {\Phi_F}\sp\,\dpr\, -\, \I\,
{\Phi'_F}\sp\,\d \nn\\[0mm]
[T\s, {\Phi_B}\sp]&=&
\I\, (d+{1\over 2})\,{\Phi_B}\sp\,\dpr\, -\, \I
\,{\Phi'_B}\sp\,\d \nn\\[0mm]
\{ T_F\s, {\Phi_F}\sp \}&=&-\,{1\over 2{\sqrt 2}}\,{\Phi_B}\sp
\,\d \nn\\[0mm]
[T_F\s, {\Phi_B}\sp]&=&
\I\, d\,{\Phi_F}\sp\,\dpr\, -\, {\I\over 2}\,
{\Phi'_F}\sp\,\d\ ,
\label{eqvir}
\eea
and likewise for ${\overline T}\s$ and ${\overline T}_F\s$.

In what follows we study superconformal deformations,
that is, variations of the stress energy superfields
\bea
T\s &\rightarrow& T\s\,+\,\delta T\s\ , \qquad T_{F}\s
\ \rightarrow\ T_{F}\s\,+\,\delta T_{F}\s \nn\\[0mm]
{\ov T}\s &\rightarrow& {\ov T}\s\,+\,\delta {\ov T}\s\ ,
\qquad {\ov T}_{F}\s\ \rightarrow\ {\ov T}_{F}\s\,+\,
\delta {\ov T}_{F}\s\ ,
\label{eqwas}
\eea
consistent with the super-Virasoro algebra.  This
requires
\bea
[\delta T\s, T\sp]\,+\,[T\s, \delta T\sp]&=&
2 \I\,\delta T\sp\,\dpr \,-\,  \I\,\delta T'\sp\,\d \nn\\[0mm]
\{ \delta T_F\s, T_F\sp \}\,+\,\{ T_F\s, \delta T_F\sp \}
&=& -\,{1\over 2\sqrt 2}\,\delta T\sp\,\d \nn\\[0mm]
[\delta T\s, T_F\sp]\,+\,[T\s, \delta T_F\sp]
&=& {3\I\over 2}\,\delta T_F\sp\,\dpr
\,-\,\I\,\delta T'_F \sp\,\d \nn\\[0mm]
[\delta T\s, {\ov T}\sp]\,+\,[T\s, \delta {\ov T}\sp]
&=& 0 \nn\\[2mm]
[\delta T\s, {\ov T}_F\sp]\,+\,[T\s, \delta {\ov T}_F\sp]
&=& 0 \nn\\[2mm]
\{\delta T_F\s, {\ov T}_F\sp \}\,+\,\{T_F\s, \delta
{\ov T}_F\sp\}&=& 0\ ,
\label{eqana}
\eea
as well as analogous conditions from
the antiholomorphic part of the algebra.

We restrict our attention to
deformations that can be written in terms of
superprimary fields.   (We relax this condition in
the Appendix.)  Therefore we make the ansatz
\be
\delta T_F\, =\, \Phi_F\ , \qquad
\quad \delta {\ov T}_F\, =\, {\tilde {\Phi}}_F\ ,
\label{eqeleu}
\ee
where $\Phi_F$ and ${\tilde{\Phi}}_F$ have dimension $(d,
\overline d)$ and $(d',\overline d')$, respectively. For
ease of notation, we suppress their dependence on the
coordinate $\sigma$.  Substituting \Ref{eqeleu} into
\Ref{eqana} and using \Ref{eqvir}, we see that $\delta
T_F$ and $\delta {\ov T}_F$ satisfy the deformation
equations provided ${\ov d} = d' = 1$, $d = {\ov d}'
={1\over 2}$, and
\be
\delta T\, =\, \delta {\ov T}\, =\, 2 \Phi_B\ ,
\ee
where $\Phi_B$ is a $(1,1)$ primary field.
These solutions are the supersymmetric
generalizations of the canonical deformations
defined in references \cite{meov}.

This formalism can be used to study
string propagation in a weak but nontrivial NS-NS
background.  We start with an undeformed theory
that describes a closed superstring in flat Minkowski
space.  The corresponding superconformal
field theory is defined by the following stress
energy superfields,
\bea
T &=& {1\over 2}\,{\eta_{\mu\nu}}{\partial}X^{\mu}
{\partial}X^{\nu}\,-\,{1\over 2}\,{\eta_{\mu\nu}}
{\psi^{\mu}}{\partial}
{\psi^{\nu}} \nn\\[0mm]
\ov{T} &=& {1\over 2}\,{\eta_{\mu\nu}}{\ov\partial}X^{\mu}
{\ov\partial}X^{\nu}\,-\,{1\over 2}\,{\eta_{\mu\nu}}
{\tilde{\psi}^{\mu}}{\ov\partial}
{\tilde{\psi}^{\nu}} \nn\\[0mm]
T_{F} &=& {1\over 2}\,{\eta}_{\mu\nu}{\psi^{\mu}}
{\partial}X^{\nu}
\nn\\[0mm]
\ov{T}_{F} &=& {1\over 2}\,{\eta}_{\mu\nu}{\tilde{\psi}^{\mu}}
{\ov\partial}X^{\nu}\ ,
\label{eqtoxou}
\eea
where $X^\mu$, $\psi^\mu$ and $\tilde{\psi}^\mu$
are world-sheet scalars and spinors, respectively.
The algebra ${\cal A}$
includes composite operators constructed out
of the matter fields\ $X^{\mu}$, $\psi^{\mu}$,
$\tilde{\psi}^{\mu}$, together with the spin fields
$S^{\alpha}$, $\tilde{S}^{\alpha}$ and the ghost
fields $b$, $c$, ${\beta}$, $\gamma$,
${\tilde b}$, ${\tilde c}$, ${\tilde{\beta}}$ and
${\tilde {\gamma}}$.  All operators are understood
to be normal ordered.

We take the deformation to be
\bea
\delta T\,=\,{\delta{\ov T}} &=& 2\Phi_B \nn\\[2mm]
&=& B_{\mu\nu}(X)
{\overline{\partial}}X^{\nu}{\partial}X^{\mu}\,
+\,{\partial_{\lambda}}
B_{\mu\nu}(X)
{\overline{\partial}}X^{\nu}{\psi^{\lambda}}
{\psi^{\mu}}\nn\\[2mm]
&&+\ \ {\partial_{\lambda}} B_{\mu\nu}(X)
{\partial}X^{\mu}{\tilde{\psi}}^{\lambda}{\tilde{\psi}}^{\nu}
\,+\,{\partial_{\rho}}{\partial_{\lambda}}B_{\mu\nu}(X)
{\psi^{\lambda}}{\psi^{\mu}}{\tilde{\psi}}^{\rho}
{\tilde{\psi}}^{\nu}\ ,
\label{eqmauro}
\eea
where $\Phi_B$ is the vertex operator for an
infinitesimal NS-NS gauge field $B_{\mu\nu}$.
The deformation is a $(1, 1)$ primary field if
\be
\Box B_{\mu\nu}(X)\,=\,0\ , \qquad
\partial^\mu B_{\mu\nu}(X)
\,=\,0\ .
\label{eqgewr}
\ee
The first of these expressions is an equation of motion
for $B_{\mu\nu}$; the second is a gauge condition.
In the Appendix, we present the deformations that
give rise to gauge-covariant equations of motion
for these fields.

The superpartners of $\delta T$ and $\delta{\ov T}$
can be found by calculating the commutators of
$\Phi_{B}$ with $T_{F}$ and ${\ov T}_F$ and demanding
that they satisfy eqs.~\Ref{eqana}.  This gives
\bea
\delta T_F &=& {1\over 2}
\bigg(B_{\mu\nu}(X){\overline{\partial}}X^{\nu}
{\psi^{\mu}}
\,+\,{\partial_{\lambda}}B_{\mu\nu}(X)
{\tilde{\psi}}^{\lambda}{\tilde{\psi}}^{\nu}
{\psi}^{\mu}\bigg)\nn\\[0mm]
\delta {\ov T}_F &=& {1\over 2}
\bigg(B_{\mu\nu}(X){\partial}X^{\mu}
{\tilde{\psi}}^{\nu}
\,+\,{\partial_{\lambda}}B_{\mu\nu}(X)
{\psi}^{\lambda}{\psi}^{\mu}{\tilde{\psi}
}^{\nu}\bigg)\ .
\label{eqgeorg}
\eea
It is tedious but straightforward
to check that $\delta T$, $\delta{\ov T}$,
$\delta T_F$ and $\delta{\ov T}_F$ satisfy the
superconformal deformation equations.  These
deformations are the same as in \cite{ovr}.

Instead of deforming the stress energy superfield,
we could have deformed the BRST charges $Q$ and
${\overline Q}$.  Nilpotency then requires
\be
\lbrace Q, \delta Q \rbrace \,=\,0\ , \quad \lbrace
{\overline Q},
\delta {\overline Q} \rbrace\,=\,0\ , \quad \lbrace Q,
\delta {\overline Q}
\rbrace\,+\,\lbrace {\overline Q}, \delta Q \rbrace
\,=\,0
\label{eqkwmh}
\ee
under the infinitesimal deformations
\be
Q \,\rightarrow\, Q\,+\,\delta Q\ , \qquad {\overline Q}
\,\rightarrow\,{\overline Q}\,+\,\delta {\overline Q}\ .
\label{eqkrokos}
\ee
The two approaches are equivalent on the local
subalgebra defined by the GSO projection.  In fact, given
a deformed BRST charge, the components of the deformed
stress energy superfield can be extracted by calculating
the commutator or anticommutator of $Q$ with the ghost field
$b$ or $\beta$.  Nilpotency of the BRST charge implies that
the deformed $T$ and $T_{F}$ obey the super-Virasoro algebra.
Conversely, given a deformed stress energy superfield, the
BRST deformations are simply
\bea
\delta Q &=&
\int d{\sigma}\bigg [c\,\delta T\,-\,{1\over 2}
{\gamma}\,\delta T_{F}\bigg]\nn\\[0mm]
\delta {\overline Q} &=& \int d{\sigma}
\bigg[{\tilde c}\,\delta {\ov T}\,-\,{1\over 2}
{\tilde{\gamma}}\,\delta {\ov T}_{F}\bigg]\ .
\label{eqpanay}
\eea
It is straightforward to verify that the deformations
\Ref{eqpanay} satisfy \Ref{eqkwmh} when
\be
Q\,=\,\int d{\sigma}\bigg[c\,\bigg(T^{(X, \psi)}\,+\,
{1\over 2} T^{(b,c, \beta, \gamma)}\bigg)
\,-\,{1\over 2}
{\gamma}\,\bigg[ T^{(X, \psi)}_{F}\,+\,{1\over 2}
T^{(b, c, \beta, \gamma)}_{F}\bigg]\bigg]\ ,
\label{eqpana0h}
\ee
and likewise for ${\ov Q}$.

\section{Spacetime Symmetries}

In string theory, the stress energy superfields ${\cal T}_\Phi
= T_{F(\Phi)} + \theta T_{\Phi}$ and ${\overline{\cal T}}_\Phi
= \Tb_{F(\Phi)} + \bar\theta \Tb_{\Phi}$ depend on the spacetime
fields $\Phi$.  Spacetime symmetries are superconformal
deformations that induce changes in the spacetime fields:
\bea
\delta T &=& \I\, [h, T_{\Phi}]
\ =\ T_{\Phi+{\delta{\Phi}}}\,-\,T_{\Phi}  \nn\\[0mm]
\delta T_F &=& \I\, [h, T_{F(\Phi)}]
\ =\ T_{F(\Phi+{\delta{\Phi}})}
\,-\,  T_{F(\Phi)} \nn\\[0mm]
\delta {\ov T} &=& \I\,[h, {\ov T}_{\Phi}]
\ =\ {\ov T}_{\Phi+{\delta{\Phi}}}\,-\,
{\ov T}_{\Phi}\nn\\[0mm]
\delta {\ov T}_F &=& \I\,[h, {\ov T}_{F(\Phi)}]
\ =\ {\ov T}_{F(\Phi+{\delta{\Phi}})}\,-\,
{\ov T}_{F(\Phi)} \ .
\label{eqlouc}
\eea
The operator $h$ is the generator of the spacetime symmetry;
it is the zero mode of a sum of dimension $(1,0)$ and $(0,1)$
currents \cite{mbo}.

Any spacetime symmetry can be described in this way, including
the gauge symmetry of a two-form field.  The generator of
two-form gauge symmetry is
\bea
h &=& {\ints}d{\theta}d{\overline{\theta}}\ [
{\xi}_{\mu}(\chi)D{\chi}^{\mu}\,-\,{\xi}_{\mu}
({\tilde{\chi}}){\overline D}{\tilde{\chi}}^{\mu}]\nn\\[0mm]
&=& \ints\left(\xi_\mu(X)\dx^\mu\,-\,\xi_\mu(X)\dbx^\mu
\,+\,{\partial_{\mu}}{\xi_{\nu}}(X)
{\psi}^{\mu}{\psi}^{\nu}
\,-\,{\partial_{\mu}}{\xi_{\nu}}(X)
{\tilde{\psi}}^{\mu}{\tilde{\psi}}^{\nu}\right)\ ,
\label{eqantaz}
\eea
where $\chi^\mu=\psi^{\mu} +{\theta}X^{\mu}$, ${\tilde{
\chi}}^\mu={\tilde{\psi}}^{\mu} +{\overline{\theta}}
X^{\mu}$ and $D$ and $\overline D$ are superspace
covariant derivatives.  The integrand is a sum of
terms of the correct dimensions provided
\be
\Box\xi^\mu(X)\,=\,0\ , \qquad
\partial_\mu\xi^\mu(X)\,=\,0\ .
\label{eqivic}
\ee

Let us check these assertions by computing the variations of the
stress energy superfields:
\bea
\I \,[h, T] &=& {\partial_{\mu}}{\xi_{\nu}}
{\overline{\partial}}X^{\mu}{\partial}X^{\nu}\,+
\,{\partial_{\lambda}}
{\partial_{\mu}}{\xi_{\nu}}
{\overline{\partial}}X^{\mu}{\psi^{\lambda}}{\psi^{\nu}}\,-\,
\partial_{\mu}{\xi_{\nu}}
{\overline{\partial}}X^{\nu}{\partial}X^{\mu}\,-
\,{\partial_{\lambda}}
{\partial_{\mu}}{\xi_{\nu}}
{\partial}X^{\mu}{\tilde{\psi}}^{\lambda}
{\tilde{\psi}}^{\nu} \nn\\
\I\,[h, T_F]&=&{1\over 2}\,{\partial_{\mu}}{\xi_{\nu}}
{\overline{\partial}}X^{\mu}{\psi^{\nu}}\,-\,
{1\over 2}\,{\partial_{\mu}}{\xi_{\nu}}
{\ov{\partial}}X^{\nu}{\psi}^{\mu}\,-\,{1\over 2}\,
{\partial_{\mu}}
{\partial_{\lambda}}{\xi_{\nu}}{\psi^{\mu}}
{\tilde{\psi}}^{\lambda}
{\tilde{\psi}}^{\nu}\,+\,{1\over 2}\,{\partial_{\mu}}
{\partial_{\lambda}}{\xi_{\nu}}{\psi^{\mu}}{\psi}^{\lambda}
{\psi}^{\nu}\nn\\
\I\,[h, {\ov T}] &=& {\partial_{\mu}}{\xi_{\nu}}
{\overline{\partial}}X^{\mu}{\partial}X^{\nu}\,+
\,{\partial_{\lambda}}
{\partial_{\mu}}{\xi_{\nu}}
{\overline{\partial}}X^{\mu}{\psi^{\lambda}}
{\psi^{\nu}}\,-\,\partial_{\mu}{\xi_{\nu}}
{\overline{\partial}}X^{\nu}{\partial}X^{\mu}\,
-\,{\partial_{\lambda}}
{\partial_{\mu}}{\xi_{\nu}}
{\partial}X^{\mu}{\tilde{\psi}}^{\lambda}
{\tilde{\psi}}^{\nu}\nn\\
\I\,[h, {\ov T}_F] &=& {1\over 2}\, {\partial_{\mu}}
{\xi_{\nu}}
{\partial}X^{\nu}{\tilde{\psi}}^{\mu}\,-\,
{1\over 2}\,{\partial_{\mu}}{\xi_{\nu}}
{\partial}X^{\mu}{\tilde{\psi}}^{\nu}\,-\,{1\over 2}\,
{\partial_{\mu}}
{\partial_{\lambda}}{\xi_{\nu}}{\tilde{\psi}}^{\mu}
{\tilde{\psi}}^{\lambda}
{\tilde{\psi}}^{\nu}\,+\,{1\over 2}\,{\partial_{\mu}}
{\partial_{\lambda}}{\xi_{\nu}}{\tilde{\psi}}^{\mu}
{\psi}^{\lambda}
{\psi}^{\nu}\ ,\nn\\
\label{eqgokic}
\eea
where, in the interest of space, we suppress the arguments
$(X)$.  From \Ref{eqmauro} and \Ref{eqgeorg} we see that
the variations can indeed be described by a $B_{\mu\nu}$
spacetime field,
\be
B_{\mu\nu}(X)\,=\,\partial_{\nu}{\xi_{\mu}}(X)
\,-\,\partial_{\mu}
{\xi_{\nu}}(X)\ .
\label{eqvuios}
\ee
The deformations \Ref{eqgokic} induce a pure-gauge
background for $B_{\mu\nu}$.  The background
preserves the gauge \Ref{eqgewr}.

This construction can be readily generalized to
an infinite class of infinitesimal gauge symmetries
\cite{mgn} and to finite symmetry transformations
(T-duality) \cite{gm}.  These higher symmetries are
generated by operators which classically have higher
dimension, such as
\be
h\,=\,\int d{\sigma}\ {\omega}_{\mu\cdots\nu\cdots
\rho\cdots\lambda}(X)\,
{\partial^{w}}X^{\mu}\cdots{\overline{\partial}
}^{u}X^{\nu}\cdots
\psi^{\rho}\cdots{\tilde{\psi}}^{\lambda}\ .
\label{eqgreolios}
\ee
The integrand is of dimension one if the
functions ${\omega}_{\mu\cdots\nu\cdots\rho\cdots\lambda
}$ satisfy differential constraints which can
be viewed as gauge conditions.  The transformation
describes a spontaneously broken spacetime symmetry because
it mixes massive and massless spacetime fields \cite{bg}.

The previous discussion can also be carried through in terms
of the BRST formalism.  Let us suppose that $Q_\Phi$ and
${\ov Q}_\Phi$ are nilpotent BRST charges, functions of
the spacetime fields, and $h$ is the zero-mode of a sum
of dimension $(1,0)$ and $(0,1)$ currents.  Then $\Phi
\ra {\Phi+{\delta{\Phi}}}$ is a spacetime symmetry if
\bea
\delta Q_\Phi &=& \I\, [ h, Q_\Phi ] \ =\
Q_{\Phi+{\delta{\Phi}}}\,-\, Q_\Phi \nn\\[0mm]
\delta {\ov Q}_\Phi &=& \I\,[ h, {\ov Q}_\Phi ]
\ =\ {\ov Q}_{\Phi+{\delta{\Phi}}} \,-\, {\ov Q}_\Phi .
\eea
For the case at hand, the BRST operator $Q$ is given
by
\bea
Q &=& {\int}d{\sigma}\, c\ \left(\,{1\over 2}
{\eta_{\mu\nu}}{\partial}X^{\mu}{\partial}X^{\nu}\,-\,
{1\over 2}{\eta_{\mu\nu}}{\psi^{\mu}}
{\partial}{\psi^{\nu}}
 \,-\, {3\over 2}{\beta}{\partial}{\gamma}\,-\,{1\over 2}
{\partial}{\beta}{\gamma}\right) \nn\\[2mm]
&&+\ {\int}d{\sigma} \ \left( bc{\partial}c
\,+\,{1\over 2}{\gamma}\,{\eta}_{\mu\nu}{\psi^{\mu}}
{\partial}X^{\nu}
\,-\,{1\over 4}b{\gamma}^{2}\right)\ ,
\label{eqasoud}
\eea
and $h$ is given in \Ref{eqantaz}.  We compute the
commutator and find
\bea
\delta Q &=& \I\,[ h, Q ]\ =\ \int d{\sigma}\ c
\ \bigg[{\partial_{\mu}}{\xi_{\nu}}
{\overline{\partial}}X^{\mu}{\partial}X^{\nu}\,+
\,{\partial_{\lambda}}
{\partial_{\mu}}{\xi_{\nu}}
{\overline{\partial}}X^{\mu}{\psi^{\lambda}}
{\psi^{\nu}}\,-\,
\partial_{\mu}{\xi_{\nu}}
{\overline{\partial}}X^{\nu}{\partial}X^{\mu}\nn\\[2mm]
&& -\ {\partial_{\lambda}}
{\partial_{\mu}}{\xi_{\nu}}
{{\partial}}X^{\mu}{\tilde{\psi}}^{\lambda}
{\tilde{\psi}}^{\nu}\bigg]
\, -\, {1\over 4}\int d{\sigma}{\gamma}\
\bigg[
{\partial_{\mu}}{\xi_{\nu}}
{\overline{\partial}}X^{\mu}{\psi^{\nu}}\,-\,
{\partial_{\mu}}{\xi_{\nu}}
{\ov{\partial}}X^{\nu}{\psi}^{\mu}\nn\\[2mm]
&& -\ {\partial_{\mu}}
{\partial_{\lambda}}{\xi_{\nu}}{\psi^{\mu}}
{\tilde{\psi}}^{\lambda}
{\tilde{\psi}}^{\nu}\,+\,{\partial_{\mu}}
{\partial_{\lambda}}{\xi_{\nu}}{\psi^{\mu}}{\psi}^{\lambda}
{\psi}^{\nu}
\bigg]\ .
\label{eqafirim}
\eea
Comparing with \Ref{eqpana0h} and \Ref{eqgokic}, we see that
this deformation can be absorbed in the two-form gauge potential
\Ref{eqvuios}.

\section{Supersymmetry Algebra}

We are now ready to compute the supersymmetry algebra in
the two-form gauge field background.  We start with the
undeformed super-Poincar\'e generators,
\bea
P^{\mu(0)}\, =\, \int d{\sigma}\, \Big({\partial X}^{\mu}
+{\ov{\partial}}X^\mu \Big)
\quad && \quad
Z^{\mu(0)}\, =\, \int d{\sigma}\, \Big({\partial X}^{\mu}
-{\ov{\partial}}X^\mu \Big)  \nn\\
Q_{\alpha}^{(-{1\over 2})}
\, =\, \int d{\sigma}\ {J}_{\alpha}^{(-{1\over 2})}
\quad && \quad {\tilde Q}_{\alpha}^{(-{1\over 2})}
\, =\, \int d{\sigma}\ {\tilde J}_{\alpha}^{(-{1\over 2})}
\label{eqakakios}
\eea
where $P^{\mu(0)}$ and $Z^{\mu(0)}$ are the spacetime translation
and winding number generators, $Q_{\alpha}^{(-{1\over 2})}$
and ${\tilde Q}_{\alpha}^{(-{1\over 2})}$ are the spacetime
supercharges, and the supersymmetry currents are given by
\be
J_{\alpha}^{(-{1\over 2})}
\, =\, S_{\alpha}e^{-{{\phi}\over 2}} \qquad\qquad
{\tilde J}_{\alpha}^{(-{1\over 2})}
\, =\, {\tilde S}_{\alpha}e^{-{
{\tilde{\phi}}\over 2}}\ .
\ee
The operators are in the canonical picture; the superscripts
indicate the ghost charges of the operators.  It is a small
calculation to show that the generators obey the following
commutation relations,
\bea
&&\qquad [ P^{\mu(0)}, P^{\nu(0)} ] \,=\, 0\ ,\qquad
[ P^{\mu(0)}, Q_{\alpha}^{(-{1\over 2})}]\, =\, 0\ ,
\qquad [ P^{\mu(0)}, {\tilde Q}_{\alpha}^{(-{1\over 2})}]
\, =\,
0\nn\\[2mm]
&&\qquad [ Z^{\mu(0)}, Z^{\nu(0)} ] \,=\, 0\ ,\qquad
[ Z^{\mu(0)}, Q_{\alpha}^{(-{1\over 2})}]\, =\, 0\ ,
\qquad [ Z^{\mu(0)}, {\tilde Q}_{\alpha}^{(-{1\over 2})}]
\, =\,
0\nn\\[4mm]
&&\{ Q_{\alpha}^{(-{1\over 2})}, Q_{\beta}^{(-{1\over 2})} \}
\, =\,  ({\gamma_{\mu}})_{\alpha\beta}\int d{\sigma}
\ {\psi}^\mu
e^{-\phi} \ ,\quad
 \{ {\tilde Q}_{\alpha}^{(-{1\over 2})},
{\tilde Q}_{\beta}^{(-{1\over 2})} \}
\, =\, ({\gamma_{\mu}})_{\alpha\beta}\int d{\sigma}
\ {\tilde{\psi}}^\mu
e^{-{\tilde\phi}} \ .\nn\\
\label{eqaswre}
\eea

To interpret this algebra, we recall the picture
changing operation that is an essential ingredient of superstring
theory \cite{she}.  Picture changing maps a BRST-invariant operator
$O^{(q)}$ of ghost charge $q$ to an equivalent operator of charge
$q+1$ via the commutator
\be
O^{(q+1)}\ =\ [Q,2{\xi}O^{(q)}]\ ,
\label{picchang}
\ee
where $\xi$ is defined through the bosonization of the superconformal
ghosts,
\be
\beta \,=\, e^{-\phi}{\partial\xi}\ , \qquad
\gamma\,=\, e^{\phi}{\eta}\ .
\label{eqariteop}
\ee
It is straightforward to show that under the picture changing,
\bea
[ Q,\,
2{\xi}{\psi}^{\mu}e^{-{\phi}}]\,+\,[\ov Q,\, {\tilde\xi}
{\tilde{\psi}}^{\mu}e^{-{\tilde{\phi}}}] &=&
{\partial X}^\mu \,+\,{\ov{\partial}}X^\mu \nn\\[2mm]
[ Q,\,
2{\xi}{\psi}^{\mu}e^{-{\phi}}]\,-\,[\ov Q,\, {\tilde\xi}
{\tilde{\psi}}^{\mu}
e^{-{\tilde{\phi}}}] &=&
{\partial X}^\mu \,-\,{\ov{\partial}}X^\mu\ .
\label{eqarchie}
\eea
This implies that
\be
P^{\mu(-1)} \,=\, \int d\sigma \, \Big(
{\psi}^{\mu}e^{-{\phi}}+{\tilde{\psi}}^{\mu}
e^{-{\tilde{\phi}}} \Big) \quad\qquad
Z^{\mu(-1)} \,=\, \int d\sigma \, \Big(
{\psi}^{\mu}e^{-{\phi}}-{\tilde{\psi}}^{\mu}
e^{-{\tilde{\phi}}} \Big)
\ee
are the momentum and winding number generators in the
$(-1)$ picture.  Using these relations, we can write the
last line of \Ref{eqaswre} in a familiar form,
\be
\{Q_{i\alpha}^{(-{1\over 2})},\, Q_{j\beta}^{(-{1\over 2})} \}
\,=\,
({\gamma^{\mu}})_{\alpha\beta}\,(\delta _{ij}P^{(-1)}_\mu
\,+\,{\epsilon}_{ij}Z^{(-1)}_\mu )\ ,
\label{eqvissi}
\ee
where $Q_{1\alpha}^{(-{1\over 2})} \equiv Q_{\alpha}^{(-{1\over 2})}
+ {\tilde Q}_{\alpha}^{(-{1\over 2})}$ and
$Q_{2\alpha}^{(-{1\over 2})} \equiv Q_{\alpha}^{(-{1\over 2})}-
{\tilde Q}_{\alpha}^{(-{1\over 2})}$.   The spacetime supercharges
close into the usual $N=2$ super-Poincar\'e algebra, modulo a
change of picture.

We now extend this analysis to the $B_{\mu\nu}$ background.
We first need to find gauge-covariant versions of the supersymmetry
generators.  We begin by computing the two-form gauge transformations
of \Ref{eqakakios},
\bea
\I\,[h, \partial X_{\mu}]&=&{1\over 2}\,[{\partial_\mu}
\xi_\nu(X)\,-\,{\partial_\nu}\xi_\mu(X)]({\partial}X^{\nu}
\,-\,{\overline\partial}X^{\nu})
\,+\,{1\over 2}\,{\partial_{\mu}}
{\partial_{\lambda}}\xi_\rho(X)({\psi}^{\lambda}{\psi}^{\rho}
\,-\,{\tilde{\psi}}^{\lambda}{\tilde{\psi}}^{\rho})\nn\\[0mm]
\I\,[h, {\ov\partial}X_{\mu}] &=& {1\over 2}\,[{\partial_\mu}
\xi_\nu(X)\,-\,{\partial_\nu}
\xi_\mu(X)]({\partial}X^{\nu}
\,-\,{\overline\partial}X^{\nu})
\,+\,{1\over 2}\,{\partial_{\mu}}
{\partial_{\lambda}}\xi_\rho(X)({\psi}^{\lambda}{\psi}^{\rho}
\,-\,{\tilde{\psi}}^{\lambda}{\tilde{\psi}}^{\rho})\nn\\[0mm]
\I\,[h, S_{\alpha}e^{-{{\phi}\over 2}}] &=&
{1\over 2}\,({\gamma}^{\rho\lambda})_{\alpha}^{\beta}
\,{\partial_\rho}\xi_\lambda(X)
\,S_{\beta}e^{-{{\phi}\over 2}}\nn\\[0mm]
\I\,[h, {\tilde S}_{\alpha}e^{-{{\tilde{\phi}}\over 2}}] &=&
\,-\,{1\over 2}\,({\gamma}^{\rho\lambda})_{\alpha}^{\beta}
\,{\partial_\rho}\xi_\lambda(X)
\,{\tilde S}_{\beta}e^{-{{\tilde{\phi}}\over 2}}\ .
\label{eqfioup}
\eea
where $h$ is given by \Ref{eqantaz}.  These expressions suggest
that we take the following operators to be the gauge-covariant
super-Poincar\'e generators in the canonical picture and the
(infinitesimal) $B_{\mu\nu}$ background:
\bea
\hat{P}^{\mu(0)}\, =\, \int d{\sigma}\, \Big(\hat{\partial} X^{\mu}
+ \hat{\ov{\partial}} X^\mu \Big)
\quad && \quad
\hat{Z}^{\mu(0)}\, =\, \int d{\sigma}\, \Big(\hat{\partial} X^{\mu}
- \hat{\ov{\partial}} X^\mu \Big) \nn\\
\hat{Q}_{\alpha}^{(-{1\over 2})}
\, =\, \int d{\sigma}\ \hat{J}_{\alpha}^{(-{1\over 2})}
\quad && \quad \hat{\tilde{Q}}_{\alpha}^{(-{1\over 2})}
\, =\, \int d{\sigma}\ \hat{\tilde{J}}_{\alpha}^{(-{1\over 2})}\ ,
\eea
where
\bea
{\hat{\partial} X}_\mu &=& {\partial X}_{\mu} \,+\,
{1\over 2}\,B_{\mu\nu}(X)\,({\partial}X^{\nu}
\,-\,{\overline\partial}X^{\nu})\,+\,{1\over 2}\,{\partial_\nu}
B_{\mu\lambda}(X)\,
({\psi^\nu}{\psi^\lambda}\,-\,{\tilde\psi}^\nu
{\tilde\psi}^\lambda) \nn\\[0mm]
{\hat{\ov{\partial}}X}_\mu &=& {\ov{\partial}}X_{\mu}\,+\,
{1\over 2}\,B_{\mu\nu}(X)\,({\partial}X^{\nu}
\,-\,{\overline\partial}X^{\nu})\,+\,{1\over 2}\,{\partial_\nu}
B_{\mu\lambda}(X)\,
({\psi^\nu}{\psi^\lambda}-{\tilde\psi}^\nu
{\tilde\psi}^\lambda)\nn\\[0mm]
{\hat J}_{\alpha}^{(-{1\over 2})} &=&
S_{\alpha}e^{-{{\phi}\over 2}}
\,+\,{1\over
4}\,({\gamma}^{\rho\lambda})_{\alpha}^{\beta}\,B_{\rho\lambda}(X)
S_{\beta}e^{-{{\phi}\over 2}}\nn\\[0mm]
{\hat{\tilde J}}_{\alpha}^{(-{1\over 2})} &=&
{\tilde S}_{\alpha}e^{-{{\tilde{\phi}}\over 2}}
\,-\,{1\over
4}\,({\gamma}^{\rho\lambda})_{\alpha}^{\beta}\,B_{\rho\lambda}(X)
{\tilde S}_{\beta}e^{-{{\tilde{\phi}}\over 2}}\ .
\label{eqagaphtos}
\eea
It is not hard to check that the generators are indeed covariant
under two-form gauge transformations.

A nontrivial $B_{\mu\nu}$ field spontaneously breaks the
translational symmetry of Minkowski space \cite{bg}.  This
can be seen from the commutator of the deformed stress energy
tensor with the gauge-covariant translation current ${\hat{\partial}}
X_{\mu}$,
\bea
[ T\s+{\delta}T\s, {\hat{\partial}} X_{\mu}\sp ]&=&
\I\,{\hat{\partial}} X_{\mu}\sp\,\dpr\,-\,\I\,{\hat{\partial}}
X_{\mu}^{\prime} \sp\,\d\nn\\[2mm]
&& -\,\I\,H_{\mu\nu\lambda}(X)\,{\partial}X^{\nu}
{\ov\partial}X^{\lambda}\,\d \nn\\[2mm]
&&-\,2\I\,{\partial_\nu} H_{\mu\rho\sigma}(X)\,
{\psi}^{\nu}{\psi}^{\rho}{\ov\partial}X^{\sigma}\,\d \nn\\[2mm]
&&-\,2\I\,{\partial_\nu}H_{\mu\rho\sigma}(X)\,
{\tilde{\psi}}^{\nu}{\tilde{\psi}}^{\rho}{\partial}X^{\sigma}
\,\d\nn\\[2mm]
&& -\,\I\,{\partial_\mu}{\partial_\lambda}
H_{\nu\rho\sigma}(X)\,{\psi}^{\nu}{\psi}^{\rho}
{\tilde{\psi}}^{\lambda}{\tilde{\psi}}^{\sigma}\,\d \ ,
\label{holovko}
\eea
where we work to first order in the $B_{\mu\nu}$ field.  The
symmetry is conserved if ${\hat{\partial}} X_{\mu}$ is primary
and of dimension one.  This requires that the field strength
$H_{\mu\nu\lambda} = 0$.  For nonzero $H_{\mu\nu\lambda}$,
the gauge-covariant translations are spontaneously broken, just
as they are for the point particle in a constant magnetic
field.

The supersymmetry currents ${\hat J}_{\alpha}^{(-{1\over
2})}$ and ${\hat{\tilde J}}_{\alpha}^{ (-{1\over 2})}$ are
also spontaneously broken for nonvanishing $H_{\mu\nu\lambda}$.
This follows from the commutator
\bea
[ T\s+{\delta}T\s, {\hat J}_{\alpha}^{(-{1\over 2})}\sp ]&=&
\I\,{\hat J}_{\alpha}^{(-{1\over 2})}\sp\,\dpr
\,-\,\I\,{\hat J}_{\alpha}^{(-{1\over 2})\prime}\sp\,\d\nn\\[2mm]
&&+\,{\I\over
2}\,H_{\mu\nu\lambda}(X)\,({\gamma^{\mu\nu}})_{\alpha}^{\beta}
\,S_{\beta}e^{-{{\phi}\over 2}}\,{\ov\partial}X^{\lambda}
\,\d\nn\\[2mm]
&&+\,\I\, ({\gamma}^{\mu\nu})_{\alpha}^{\beta}\,{\partial_\rho}
H_{\mu\nu\lambda}\,S_{\beta}e^{-{{\phi}\over 2}}\,
{\tilde{\psi}}^{\rho}{\tilde{\psi}}^{\lambda}\, \d\ .
\label{housin}
\eea
A nonzero field strength spontaneously breaks spacetime
supersymmetry, as expected from supergravity.

Even though the super-Poincar\'e symmetries are spontaneously
broken, one can still compute the supersymmetry algebra
in the $B_{\mu\nu}$ background.  It is a small exercise
to show that the gauge-covariant supercharges obey the
following anticommutation relations
\bea
\{ {\hat Q}_{\alpha}^{(-{1\over 2})},
{\hat Q}_{\beta}^{(-{1\over 2})} \}
&=& ({\gamma^{\mu}})_{\alpha\beta}\int
d{\sigma}\bigg[{\psi}_{\mu}e^{-{\phi}}
\,+\,{1\over 2}\,B_{\mu\nu}(X)
{\psi}^{\nu}e^{-{\phi}}\bigg] \nn\\[0mm]
\{ {\hat{\tilde Q}}_{\alpha}^{(-{1\over 2})},
{\hat{\tilde Q}}_{\beta}^{(-{1\over 2})} \}
&=& ({\gamma^{\mu}})_{\alpha\beta}\int
d{\sigma}\bigg[{\tilde{\psi}}_{\mu}
e^{-{\tilde{\phi}}}
\,-\,{1\over 2}\,B_{\mu\nu}(X)
{\tilde{\psi}}^{\nu}e^{-{\tilde{\phi}}}\bigg]
\nn\\[0mm]
\{ {\hat Q}_{\alpha}^{(-{1\over 2})},
{\hat{\tilde Q}}_{\beta}^{(-{1\over 2})} \} &=& 0\ ,
\label{eqpyrs}
\eea
to first order in the $B_{\mu\nu}$ field.  Rewriting these
expressions in terms of ${\hat Q}_{1\alpha}^{(-{1\over 2})}$
and ${\hat Q}_{2\alpha}^{(-{1\over 2})}$, we find
\bea
\{ {\hat Q}_{i\alpha}^{(-{1\over 2})},
{\hat Q}_{j\beta}^{(-{1\over 2})}\} &=& \delta _{ij}\,
({\gamma^{\mu}})_{\alpha\beta}\int d{\sigma}\bigg[{\psi}_{\mu}e^{-{\phi}}
\,+\,{\tilde{\psi}}_{\mu}
e^{-{\tilde{\phi}}}
\,+\,{1\over 2}\,B_{\mu\nu}(X)\,({\psi}^{\nu}e^{-{\phi}}
\,-\,{\tilde{\psi}}^{\rho}e^{-{\tilde{\phi}}})\bigg] \nn\\[0mm]
&+&{\epsilon}_{ij}\,
({\gamma^{\mu}})_{\alpha\beta}\int d{\sigma}\bigg[{\psi}_{\mu}e^{-{\phi}}
\,-\,{\tilde{\psi}}_{\mu}
e^{-{\tilde{\phi}}}
\,+\,{1\over 2}\,B_{\mu\nu}(X)\,({\psi}^{\nu}e^{-{\phi}}
\,+\,{\tilde{\psi}}^{\rho}e^{-{\tilde{\phi}}})\bigg]\nn\\
\label{eqasvopl}
\eea

To interpret this expression, we must define the picture
changing operation in the gauge field background.
{}From \Ref{picchang} we find
\be
\delta O^{(q+1)}\ =\ [\delta Q, 2{\xi}O^{(q)}]\,+\,
[Q, 2\delta {\xi}O^{(q)}]\,+\,[Q, 2{\xi}\delta O^{(q)}]\ ,
\label{eqavionm}
\ee
where $\delta O^{(p)}$ is the deformation in the $p$
picture, $\delta {\xi}$ is the deformation of the ghost $\xi$,
and $\delta Q$ is the deformation of the BRST charge.
Using this relation, it is not hard to show that the first
term on the right hand side of \Ref{eqasvopl} is the
gauge-covariant momentum generator in the $-1$ picture,
\bea
\int d\sigma \,\bigg\{\bigg[ Q+\delta Q,
2{\xi}\bigg({\psi}_{\mu}e^{-{\phi}}
\,+\,{1\over 2}\,B_{\mu\nu}(X){\psi}^{\nu}e^{-{\phi}}\bigg)
\bigg]\ + &&\nn\\[3mm]
\bigg[ {\ov Q}+\delta {\ov Q},2{\tilde\xi}
\bigg({\tilde{\psi}}_{\mu}
e^{-{\tilde{\phi}}}
\,+\,{1\over 2}B_{\mu\nu}(X)\,
{\tilde{\psi}}^{\nu}e^{-{\tilde{\phi}}}\bigg)\bigg]\bigg\}
&=& \int d\sigma \,\Big(
{\hat{\partial} X}_\mu\,+\,{\hat{\ov{\partial}}X}_\mu \Big) \ .
\eea
The second term is the gauge-covariant winding number generator
in the same picture.  Combining these results, we find
\be
\{ {\hat Q}_{i\alpha}^{(-{1\over 2})},
{\hat Q}_{j\beta}^{(-{1\over 2})}\} \ =\
({\gamma^{\mu}})_{\alpha\beta}\ ( \delta _{ij}\,\hat{P}^{(-1)}_\mu
\,+\, \epsilon_{ij}\,\hat{Z}^{(-1)}_\mu )\ .
\ee

We can use similar techniques to compute the remaining parts
of the supersymmetry algebra:
\bea
[ {\hat P}_{\mu}^{(0)}, {\hat P}_{\nu}^{(0)} ] &=&
- \int d{\sigma}\ \Big[ H_{\mu\nu\lambda}(X)
\,({\partial}X^{\lambda}
\,-\,{\overline\partial}X^{\lambda})\,+\,
{\partial}_{\lambda}H_{\mu
\nu\rho}(X)({\psi^{\lambda}}{\psi^{\rho}}\,-\,
{\tilde{\psi}}^{\lambda}{\tilde{\psi}}^{\rho}) \Big] \nn\\[0mm]
[ {\hat Q}_{\alpha}^{(-{1\over 2})}, {\hat P}_{\mu}^{(0)} ] &=&
({\gamma}^{\rho\lambda})_{\alpha}^{\beta}\int d{\sigma}
\ H_{\mu\rho\lambda}(X)\,
S_{\beta}e^{-{{\phi}\over 2}}\nn\\[0mm]
[ {\hat{\tilde Q}}_{\alpha}^{(-{1\over 2})}, {\hat P}_{\mu}^{(0)} ]
&=& -({\gamma}^{\rho\lambda})_{\alpha}^{\beta}\int d{\sigma}
\ H_{\mu\rho\lambda}(X)\,
{\tilde S}_{\beta}e^{-{{\tilde{\phi}}\over 2}}\ .
\label{eqxryso}
\eea
We see that the gauge-covariant momenta and supersymmetry
charges cease to commute in the presence of a non-trivial
NS-NS background field.  For the case of constant $H_{\mu\nu\lambda}$,
however, the commutators \Ref{eqxryso} simplify considerably.
We find
\bea
[ {\hat P}_{\mu}^{(0)}, {\hat P}_{\nu}^{(0)} ] &=&
-\,H_{\mu\nu\lambda}\,Z^{\lambda(0)}\nn\\[0mm]
[ {\hat Q}_{\alpha}^{(-{1\over 2})}, {\hat P}_{\mu}^{(0)} ] &=&
({\gamma}^{\rho\lambda})_{\alpha}^{\beta}
\ H_{\mu\rho\lambda}\,{\hat Q}_{\beta}^{(-{1\over 2})}\nn\\[0mm]
[ {\hat{\tilde Q}}_{\alpha}^{(-{1\over 2})}, {\hat P}_{\mu}^{(0)} ]
&=& -\,({\gamma}^{\rho\lambda})_{\alpha}^{\beta}
\ H_{\mu\rho\lambda}\,
{\hat{\tilde Q}}_{\beta}^{(-{1\over 2})}\ .
\label{eqxrdo}
\eea
This algebra is similar to that of the supersymmetric point
particle in a constant electromagnetic background.

We have checked our results by verifying that the Jacobi identity
still holds.  For example, we compute
\bea
&&[ {\hat Q}_{\alpha}^{(-{1\over 2})},
[ {\hat P}_{\mu}^{(0)},
{\hat P}_{\nu}^{(0)} ]]\,+\,
[ {\hat P}_{\mu}^{(0)}, [{\hat P}_{\nu}^{(0)},
{\hat Q}_{\alpha}^{(-{1\over 2})}]]\,+\,[{\hat P}_{\nu}^{(0)},
[{\hat Q}_{\alpha}^{(-{1\over 2})},
{\hat P}_{\mu}^{(0)}]]\nn\\[0mm]&&\qquad\qquad
=\ ({\gamma}^{\lambda\rho})_{\alpha\beta}\int d{\sigma}\
({\partial_{\lambda}}H_{\mu\nu\rho}\,-\,{\partial_{\rho}}
H_{\lambda\mu\nu}\,-\,{\partial_{\mu}}
H_{\nu\rho\lambda}\,+\,{\partial_{\nu}}
H_{\rho\lambda\mu})\ ,
\label{eqjhop}
\eea
which vanishes because of the Bianchi identity.

A straightforward calculation of the commutator of the
string coordinate $X^{\mu}$ with the generator of two-form
gauge transformations shows that the string coordinate
does not change with a $B_{\mu\nu}$ background
field. This is in contrast to the open string, in
which case the string coordinate deforms and becomes
non-commutative \cite{noncomu}.

\section{Magnetic Supersymmetry}

In the previous section we saw that a $B_{\mu\nu}$
field spontaneously breaks spacetime supersymmetry.
It is interesting to ask whether any deformations of
the spacetime symmetries remain conserved in this
background.  In this section we shall see that there
are indeed such generators for constant $H_{\mu\nu\lambda}$.
We call them ``magnetic" super-Poincar\'e generators in
analogy to the magnetic translation operators that can
be constructed for point particles in a constant magnetic
field \cite{hall}.

The basic approach is as before.  We start by deforming the
gauge-covariant super-Poincar\'e currents \Ref{eqagaphtos}
by a sum of $(0,1)$ and $(1,0)$ operators.  We then compute
the conditions that follow from the requirement that the
new currents be primary and dimension one with respect
to the deformed stress tensor.  We find that these conditions
require $H_{\mu\nu\lambda}$ to be constant, and furthermore,
that the deformed currents be of the following form:
\bea
{\partial} X^M_{\mu} &=& {\hat{\partial} X}_{\mu}
\,-\,H_{\mu\nu\lambda}\,X^\lambda \partial X^\nu
\,-\,H_{\mu\nu\lambda}\,\psi^\nu \psi^\lambda\nn\\[0mm]
{\ov{\partial}} X^M_{\mu}  &=& {\hat{\ov{\partial}} X}_{\mu}
\,+\,H_{\mu\nu\lambda}\,X^\lambda \ov{\partial} X^\nu
\,+\,H_{\mu\nu\lambda}\,\tilde{\psi}^\nu \tilde{\psi}^\lambda\nn\\[0mm]
{ J}_{\alpha}^{(-{1\over 2})M} &=&
{\hat J}_{\alpha}^{(-{1\over 2})}
\,-\,{1\over2}\,({\gamma}^{\mu\nu})_{\alpha}^{\beta}\,H_{\mu\nu\lambda}\,
X^\lambda \, S_{\beta}e^{-{\phi \over 2}}\nn\\[0mm]
{\tilde{J}}_{\alpha}^{(-{1\over 2})M} &=&
\hat {\tilde{J}}_{\alpha}^{(-{1\over 2})}
\,+\,{1\over2}\,({\gamma}^{\mu\nu})_{\alpha}^{\beta}\,H_{\mu\nu\lambda}\,
X^\lambda \, S_{\beta}e^{-{ \phi\over 2}}\ .
\label{eqagaphtos2}
\eea
The index $M$ indicates that these are conserved, gauge-covariant,
``magnetic" super-Poincar\'e currents in a constant
$H_{\mu\nu\lambda}$ background.

Once we have the magnetic currents, it is a simple exercise to
compute the magnetic supersymmetry algebra.  We find
\be
\{ { Q}_{i\alpha}^{(-{1\over 2})M},
{ Q}_{j\beta}^{(-{1\over 2})M}\} \ =\
({\gamma^{\mu}})_{\alpha\beta}\ ( \delta _{ij}\,{P}^{(-1)M}_\mu
\,+\, \epsilon_{ij}\,{Z}^{(-1)M}_\mu )\ ,
\label{eqxrdo3}
\ee
and
\bea
[ { P}_{\mu}^{(0)M}, { P}_{\nu}^{(0)M} ] &=&
2\,H_{\mu\nu\lambda}\,Z^{\lambda(0)M}\nn\\[0mm]
[ { Q}_{\alpha}^{(-{1\over 2})M}, { P}_{\mu}^{(0)M} ] &=&
- ({\gamma}^{\rho\lambda})_{\alpha}^{\beta}
\ H_{\mu\rho\lambda}\,{ Q}_{\beta}^{(-{1\over 2})M}\nn\\[0mm]
[ {{\tilde Q}}_{\alpha}^{(-{1\over 2})M}, { P}_{\mu}^{(0)M} ]
&=& ({\gamma}^{\rho\lambda})_{\alpha}^{\beta}
\ H_{\mu\rho\lambda}\,
{{\tilde Q}}_{\beta}^{(-{1\over 2})M}\ .
\label{eqxrdo2}
\eea
In these expressions,
\bea
{P}_{\mu}^{(0)M} &=& \int d\sigma \,\Big(
{\partial} X^M_{\mu} \,+ \, {\ov{\partial}} X^M_{\mu} \Big)
\nn\\
{Z}_{\mu}^{(0)M} &=& \int d\sigma \,\Big(
{\partial} X^M_{\mu} \,- \, {\ov{\partial}} X^M_{\mu} \Big)
\eea
and
\bea
{P}^{(-1)M}_\mu &=& {\hat{P}}^{(-1)}_\mu \,-\, H_{\mu\nu\lambda}\,
\int d\sigma\,X^\lambda\,(\psi^\nu e^{-\phi} - \tilde{\psi}^\nu  e^{-\tilde
\phi}) \nn\\
{Z}^{(-1)M}_\mu &=& {\hat{Z}}^{(-1)}_\mu \,-\, H_{\mu\nu\lambda}\,
\int d\sigma\,X^\lambda\,(\psi^\nu e^{-\phi} + \tilde{\psi}^\nu  e^{-\tilde
\phi})
\eea
are the magnetic translation and winding generators in the $0$ and
$-1$ pictures.  This is the magnetic supersymmetry algebra that
holds in a constant $H_{\mu\nu\lambda}$ background.

\section{Conclusions}

In this paper we discussed deformations of the fermionic
string.  We showed how to deform the stress tensor, the
supercurrent and the BRST charges in a way consistent
with superconformal invariance.  We used the technique
to study superstring propagation in a nontrivial two-form
NS-NS background.

Our main result was the construction of the gauge-covariant
super-Poincar\'e generators in the presence of a $B_{\mu\nu}$
field.  We found that the $B_{\mu\nu}$ field generically
breaks spacetime supersymmetry.  For the case of constant
field strength $H_{\mu\nu\lambda}$, we found ``magnetic"
extensions of the spacetime super-Poincar\'e generators.
The magnetic generators are conserved and gauge covariant;
they are generalizations of the magnetic
translation operators that can be constructed for point
particles in a constant magnetic field.  For the case
at hand, the magnetic super-Poincar\'e generators
close into a magnetic extension of the spacetime
supersymmetry algebra.

The techniques presented here can be readily extended
to the case of a weak Ramond-Ramond background.  Work
along these lines is currently in progress.

\vspace{0.25in}
I.G.~would like to thank C.~Bering, J.~Liu, B.~Morariu,
V.~Nair and A.~Polychronakos for useful discussions.
This work was supported in part by the Department of Energy,
contract number DE-FG02-91ER40651-TASKB, and the
National Science Foundation, grant NSF-PHY-9970781.

\section*{Appendix}

In this appendix we show how to describe a NS-NS
two-form potential in an arbitrary gauge \cite{mg}.
We do this by first performing an arbitrary gauge
transformation about flat spacetime. We then replace
the transformation parameters by the
$B_{\mu\nu}$ field.

We start by computing the commutator of $h$, the generator
of a symmetry transformation, with the stress tensor $T$
and the supercurrent $T_F$. We assume that $h$ has
the form \Ref{eqantaz}, but we do not impose the
differential constraints on $\xi_{\mu}(X)$. This gives
\bea
\I \, [h, T]&=&
\partial_\mu\xi_\nu\,{\overline{\partial}}
X^{\mu}{\partial}X^{\nu}\,+\,{\partial_{\lambda}}
{\partial_{\mu}}{\xi_{\nu}}\,{\overline{\partial}}
X^{\mu}{\psi^{\lambda}}{\psi^{\nu}}\,-\,{1\over 2}\,
\Box\partial_\mu\xi_\nu\,{\overline{\partial}}
X^{\mu}{\partial}X^{\nu}\nn\\[2mm]&&\,-\,{1\over 2}\,
\Box{\partial_{\lambda}}{\partial_{\mu}}{\xi_{\nu}}\,
{\overline{\partial}}X^{\mu}{\psi^{\lambda}}{\psi^{\nu}}
\,+\,\partial_\nu\partial_\lambda\partial_\mu\xi^\mu\,
\dx^\nu\dbx^\lambda\,-\,{1\over 2}\,\partial_\nu
\partial_\lambda\partial_\mu\xi^\mu\,\dx^\nu
\dx^\lambda\nn\\[2mm]&&\,-\,{1\over 2}\,\partial_\nu
\partial_\mu\xi^\mu\,{\partial}^2 X^{\nu}\,-\,{1\over 2}\,
\partial_\nu\partial_\lambda\partial_\mu\xi^\mu\,\dbx^\nu
\dbx^\lambda\,-\,{1\over 2}\,\partial_\nu\partial_\mu\xi^\mu\,
{\overline{\partial}}^2X^{\nu}\nn\\[2mm]&&\,+\,{1\over 2}\,
\Box\partial_\mu\xi_\nu\,\dx^\mu\dx^\nu\,+\,{1\over 2}\,
\Box\xi_\mu\,{\partial^2}X^{\mu}\,+\,{1\over 2}\,
\Box\partial_\nu\partial_\mu\xi_\lambda
\,\dbx^\nu
\psi^\mu\psi^\lambda\nn\\[2mm]&&\,-\,{1\over 2}\,
\Box\partial_\nu\partial_\mu\xi_\lambda\,\dx^\nu\psi^\mu
\psi^\lambda\,+\,{1\over 2}\,\Box\partial_\mu\xi_\nu\,
\partial\psi^\mu\psi^\nu\,+\,{1\over 2}\,\Box\partial_\mu\xi_\nu\,
\psi^\mu\partial\psi^\nu\nn\\[2mm]&&\,-\,{1\over 2}\,
\Box\partial_\nu\xi_\mu\,{\overline{\partial}}X^{\mu}
{\partial}X^{\nu}\,+\,{1\over 2}\,\Box\partial_\nu
\xi_\mu\,{\overline{\partial}}X^{\mu}
{\overline{\partial}}X^{\nu}\,-\,\partial_\nu\xi_\mu\,
{\overline{\partial}}X^{\mu}
{\partial}X^{\nu}\nn\\[2mm]&&\,-\,{1\over 2}\,
\Box\partial_\nu\partial_\mu\xi_\lambda\,
\dx^\nu
{\tilde\psi}^\mu{\tilde\psi}^\lambda\,-{\partial_\lambda}
{\partial_\mu}\xi_\nu{\partial}X^\mu{\tilde\psi}^\lambda
{\tilde\psi}^\nu\, +\,{1\over 2}\,
\Box\partial_\nu\partial_\mu\xi_\lambda\,\dbx^\nu
{\tilde\psi}^\mu{\tilde\psi}^\lambda\nn\\[2mm]&&\,+\,{1\over 2}\,
\Box\xi_\mu\,{\overline{\partial}}^2X^{\mu}\,
-\,{1\over 2}\,\Box\partial_\mu\xi_\nu\,{\tilde\psi}^\mu
{\ov\partial}{\tilde\psi}^\nu\,-\,{1\over 2}\,\Box\partial_\mu
\xi_\nu\,{\ov\partial}{\tilde\psi}^\mu{\tilde\psi}^\nu
\label{eqristons}
\eea
and
\bea
\I\,
[h, T_F] &=&  {1\over 2}\,{\partial_{\mu}}{\xi_{\nu}}\,
{\overline{\partial}}X^{\mu}{\psi^{\nu}}\,+\,{1\over 2}\,
\Box\partial_\mu\xi_\nu\,{\partial X}^{\mu}\psi^\nu\,-\,
{1\over 2}\,\Box\partial_\mu\xi_\nu\,{\overline{\partial}}
X^{\mu}\psi^\nu \,+\,{1\over 2}\,
\Box\xi_{\mu}\,{\partial}{\psi}^{\mu}\nn\\[2mm]&&\,-\,{1\over 2}\,
{\partial_\mu}{\partial_\nu}{\partial_{\lambda}}\xi^\mu\,
{\partial}X^{\nu}{\psi^{\lambda}}\,+\,{1\over 2}{\partial_\mu}
{\partial_\nu}{\partial_{\lambda}}\xi^\mu\,{\overline{\partial}}
X^{\nu}{\psi^{\lambda}}\,-\,{1\over 2}\,{\partial_\mu}
{\partial_{\nu}}\xi^\mu\,{\partial}{\psi^{\nu}}\nn\\[2mm]&&\,+\,
{1\over 2}\,{\partial_\mu}{\partial_\lambda}\xi_\nu\,{\psi^\mu}
{\psi^\lambda}{\psi^\nu}\,-\,{1\over 2}\,{\partial_{\mu}}
{\xi_{\nu}}\,{\overline{\partial}}X^{\nu}{\psi^{\mu}}\,-\,
{1\over 2}\,{\partial_\mu}{\partial_\lambda}\xi_\nu\,
{\psi^\mu}{\tilde{\psi}}^\lambda{\tilde{\psi}}^\nu
\label{eqwsana}
\eea
Substituting $B_{\mu\nu} =\partial_\mu\xi_\nu -\partial_\nu
\xi_\mu$ into these equations, we find 
the deformations that describe two-form propagation in
flat spacetime,
\bea
\delta T&=&\Big(B_{\mu\nu}(X)\,-\,{1\over 2}\,
\Box B_{\mu\nu}(X)\,-\,{\partial_\nu}{\partial^\lambda}
B_{\lambda\mu}(X)\Big)\,{\overline{\partial}}X^{\mu}
{\partial}X^{\nu}\nn\\[2mm]&&\,-\,\Bigg({1\over 2}\,
\Box{B}_{\mu\nu}\,-\,{1\over 2}{\partial_\mu}{\partial^\lambda}
B_{\lambda\nu}(X)\Bigg)\,\dx^\mu\dx^\nu\,+\,{1\over 2}\,
\partial_\mu\partial^{\lambda}B_{\lambda\nu}(X)\,
\dbx^\mu\dbx^\nu\nn\\[2mm]&&\,+\,{1\over 2}\,
{\partial^\mu}B_{\mu\nu}(X)\,{\partial}^2
X^{\nu}\,+\,{1\over 2}\,{\partial^\mu}B_{\mu\nu}(X)
\,{\overline{\partial}}^2X^{\nu}\nn\\[2mm]&&\,+\,
{1\over 2}\,\Box\partial_{\lambda}B_{\nu\mu}(X)\,\dx^\mu
\psi^\lambda\psi^\nu\,-\,\Bigg({\partial_{\lambda}}B_{\nu\mu}
(X)\,-\,{1\over 2}\,\Box\partial_{\lambda}{B}_{\nu\mu}(X)
\Bigg)\,{\overline{\partial}}X^{\mu}{\psi^{\lambda}}
{\psi^{\nu}}\nn\\[2mm]&&\,-\,\Bigg({\partial_{\lambda}}
B_{\mu\nu}(X)\,+\,{1\over 2}\,\Box\partial_{\lambda}
{B}_{\mu\nu}(X)\Bigg)\,{\partial}X^{\mu}{\tilde{\psi}}^{\lambda}
{\tilde{\psi}}^{\nu}\,+\,{1\over 2}\,\Box\partial_{\lambda}
{B}_{\mu\nu}(X)\,{\ov{\partial}}X^{\mu}{\tilde{\psi}}^{\lambda}
{\tilde{\psi}}^{\nu}\nn\\[2mm]&&\,-\,{1\over 2}\,
\Bigg({\partial_\lambda}{\partial^\mu}B_{\nu\mu}(X)\,-\,
{\partial_\nu}{\partial^\mu}B_{\lambda\mu}(X)\Bigg)\,{\partial}
{\psi}^\lambda{\psi}^\nu\,-\,{\partial_{\rho}}
{\partial_{\lambda}}B_{\mu\nu}(X)\,{\psi^{\lambda}}{\psi^{\mu}}
{\tilde{\psi}}^{\rho}
{\tilde{\psi}}^{\nu}\nn\\[2mm]&&\,-\,{1\over 2}\,\Bigg({\partial_\lambda}
{\partial^\mu}B_{\mu\nu}(X)\,-\,{\partial_\nu}{\partial^\mu}
B_{\mu\lambda}(X)\Bigg)\,{\ov{\partial}}{\tilde{\psi}}^\lambda
{\tilde{\psi}}^\nu
\label{eqasynd}
\eea
and
\bea
\delta T_F &=& \Bigg(
{1\over 2}\,B_{\mu\nu}(X)\,-\,{1\over 2}\,\Box B_{\mu\nu}(X)
\,-\,{1\over 2}\,{\partial_\nu}{\partial^\lambda}
B_{\lambda\mu}(X)\Bigg)\,{\overline{\partial}}X^{\mu}
{\psi^{\nu}}\nn\\[2mm]&&\,-\,\Bigg({1\over 2}\,
\Box B_{\mu\nu}(X)\,-\,{1\over 2}\,{\partial_\mu}
{\partial^\lambda}B_{\lambda\nu}(X)\Bigg)\,{\partial}
X^{\nu}{\psi^{\mu}}\nn\\[2mm]&&\,-\,{1\over 2}\,{\partial^\mu}
B_{\nu\mu}(X)\,{\partial}{\psi^{\nu}}\,-\,
{1\over 2}\,{\partial_\lambda}B_{\mu\nu}(X)\,{\psi^\mu}
{\tilde{\psi}}^{\nu}{\tilde{\psi}}^{\lambda}
\eea
One can check that
the deformations ${\delta T}$ and $\delta T_F$ satisfy the
super-Virasoro algebra provided
\be
\Box{B}_{\mu\nu}\,-\,{\partial_{\mu}}
{\partial^{\lambda}}B_{\lambda\nu}\,-\,{\partial_{\nu}}
{\partial^{\lambda}}B_{\mu\lambda}\,=\,0\ .
\ee
This is nothing but the gauge-covariant equation of motion
for $B_{\mu\nu}$.  Imposing the gauge condition
$\partial^\mu B_{\mu\nu}=0$, we recover the gauge-fixed
expressions discussed in the previous sections.

\end{document}